# Strain-adjustable Reflectivity of Polyurethane Nanofiber Membrane for Thermal Management Applications

Xin Li[a], Zhenmin Ding[a], Giuseppe Emanuele Lio[b,c], Jiupeng Zhao[a], Hongbo Xu[a, *], Lorenzo Pattelli [b,d *], Lei Pan[a, *], Yao Li[e, *]

[a] School of Chemistry and Chemical Engineering, Harbin Institute of Technology, Harbin, 150001, PR China

[b] European Laboratory for Non-Linear Spectroscopy (LENS), Sesto Fiorentino, 50019, Italy

[c] Department of Physics and Astronomy, University of Florence, Sesto Fiorentino, 50019, Italy

[d] Istituto Nazionale di Ricerca Metrologica (INRiM), Turin, 10135, Italy

[e] Center for Composite Materials and Structure, Harbin Institute of Technology, Harbin, 150001, China

*Corresponding author: iamxhb@hit.edu.cn, l.pattelli@inrim.it, panlei@hit.edu.cn, liyao@hit.edu.cn

**Abstract**

Passive radiative cooling technologies are highly attractive in pursuing sustainable development. However, current cooling materials are often static, which makes it difficult to cope with the varying needs of all-weather thermal comfort management. Herein, a strategy is designed to obtain flexible thermoplastic polyurethane nanofiber (Es-TPU) membranes via electrospinning, realizing reversible in-situ solvent-free switching between radiative cooling and solar heating through changes in its optical reflectivity by stretching. In its radiative cooling state (0% strain), the Es-TPU membrane shows a high and angular-independent reflectance of 95.6% in the 0.25-2.5 μm wavelength range and an infrared emissivity of 93.3% in the atmospheric transparency window (8-13 μm), reaching a temperature drop of 10 °C at midday, with a corresponding cooling power of 118.25 W/m$^2$. The excellent mechanical properties of the Es-TPU membrane allows the continuous adjustment of reflectivity by reversibly stretching it, reaching a reflectivity of 61.1% (ΔR=34.5%) under an elongation strain of 80%, leading to a net temperature increase of 9.5 °C above ambient of an absorbing substrate and an equivalent power of 220.34 W m$^{-2}$ in this solar heating mode. The strong haze, hydrophobicity and outstanding aging resistance exhibited by this scalable membrane hold promise for achieving uniform illumination with tunable strength and efficient thermal management in practical applications.

**Keywords**: Passive radiative cooling, Electrospinning, TPU nanofiber membrane, Adjustable reflectivity, Mechanical strain, Thermal management

## 1. Introduction

Passive radiative cooling (PRC) has proven to be a promising electricity-free cooling strategy for dissipating heat of terrestrial objects through the atmospheric transparency window (8-13 μm) [1-5], and have attracted much attention for its potential of mitigating global warming [6] and the urban heat island effect [7, 8]. Sub-ambient daytime cooling applications under direct sunlight illumination is particularly attractive and challenging to achieve [9-11]. In fact, this requires materials that are engineered to exhibit high reflectivity (>95%) over the whole solar spectrum, combined with a high thermal emission capacity in the atmospheric transparency window [12, 13]. Even when these conditions are met, however, most PRC systems proposed up to now in the literature are static, that is, the cooling capacity of the photonic structure is fixed after fabrication [14]. While it makes sense to seek a high cooling power during the hot weather [15], the opposite is true during the cold season, where enhanced daylighting and even harnessing heat from the sun for thermal heating would be desirable instead [16-20]. Therefore, there is strong need for a practical radiative cooling system comprising dynamically responsive elements allowing a continuous tuning of its properties for thermal comfort modulation [21-23].

To date, switchable technologies capable of both radiative cooling and solar heating in a single system have been preliminarily explored. Janus films based on radiative thermal management were used in thermal comfort regulation of buildings and clothing [24-28]. Extensive studies on Janus cooling/heating switching films have focused on mechanically bonded multi-layer structures with different photothermal and radiative cooling properties [29]. This approach, however, required to manually flip the film to obtain the cooling-heating switch, which is difficult for large area building materials and incompatible with in-situ switching. A dual-mode device (parallel Janus dual function layer) with electrostatically-controlled thermal contact conductance fabricated by Li et al. [30] showed cooling and heating power densities of 71.6 W m$^{-2}$ and 643.4 W m$^{-2}$, respectively-yet still requiring the superposition of two or more functional layers, which complicated the preparation process. Alternative approaches have been explored seeking materials that can tune their solar transmittance under external stimuli. Recent results in this direction include for instance a thermochromic poly(N-isopropylacrylmide) (PNIPAm) hydrogel which exhibited a broad visible light reflectance modulation range at varying ambient temperatures [31-33]. A polyvinylidene fluoride (PVDF) film-encapsulated

PNIPAm composite prepared by Wu et al. was used to realize temperature control over a cooling-heating temperature range of 9.5 °C [34]. The same passive response to external temperature had also been applied to smart hydrogels to use them as temperature detectors with a visual response [35]. Still, this mechanism did not offer control over the desired cooling or heating power in order to adjust it in accordance to personal thermal comfort needs. Another strategy that has been explored by several authors involved the infiltration of liquids matching the refractive index of porous coating films (such as isopropanol (IPA), water, ethylene glycol, etc.) to achieve in-situ control of their optical reflectivity. This approach could be realized both in a passive fashion by exploiting the vapor-to-liquid transition in the porous structure induced by external temperature difference [36, 37], or actively by reversible wetting of the porous polymer coating with alcohol or water, such as in the case of poly(vinylidene fluoride-co-hexafluoropropene) (P(VdF-HFP)) and polytetrafluoroethene (PTFE) [38]. Although this method allowed in-situ regulation, the presence and circulation of liquids inside the cooling materials represented a significant complication, imposing also the presence of liquid-tight containers to encapsulate the switchable coating.

Moreover, most of these previous examples realize only a switching mechanism between two discrete cooling/heating states, with no intermediate modulation. Therefore, a continuously adjustable polymer membrane whose high reflectivity in the solar wavelength could be tuned mechanically would be highly desirable. In order to obtain effective PRC materials with high solar reflectance, light scattering interactions should be carefully designed over a broad wavelength range [3, 39, 40]. A particularly attractive method to fabricate highly reflective materials over large areas is that of electrospinning, due to the good control that can be obtained on the diameter range and areal density of the spinned nanofibers [41-44]. Although PDMS is an excellent elastomer with high infrared emissivity, it is not suitable for electrospinning due to its low surface energy [45-47]. Thermoplastic polyurethane (TPU), on the other hand, is an environmentally friendly elastomer with high strength, good toughness, shear and abrasion resistance, and is the best choice for the preparation of radiative cooling materials with tensile properties [48, 49].

Herein, we developed an in-situ solvent-free tuning mechanism between radiative cooling and solar heating of a nanofiber membrane, based on the reversible adjustment of its reflectivity by mechanical deformation obtained, e.g., via manual or motorized stretching. Specifically, the initial Es-TPU membrane (Es-TPU-0%) exhibited a strong reflectivity of 95.6% over the solar spectrum (0.25-2.5 μm wavelength) and infrared emissivity of 93.3% in the atmospheric transparency window (8-13 μm wavelength), reaching a cooling power of 118.25 W m$^{-2}$. When stretched to a strain of 80% (Es-TPU-80%), the measured solar reflectivity decreased to 61.1%. As a result, the membrane showed a maximum daytime temperature drop of 10 °C in its high-reflection configuration or a solar heating effect of 9.5 °C in its low-reflection state. The transition between the two configurations was completely reversible and showed excellent cycle stability, offering dynamic thermal comfort adjustment based on the subjective personal needs of individuals, with considerable application prospects in all-weather energy saving applications.

## 2. Experimental section

*2.1 Materials*

Thermoplastic Urethane (TPU, 8795A) was obtained from Bayer Co., Ltd. Germany. N, N-dimethylformamide (DMF) was purchased from Sinopharm Chemical Reagent Co., Ltd. China. All purchased reagents were directly used without further purification.

*2.2 Fabrication of the TPU Nanofibrous Membranes.*

TPU nanofiber membrane was obtained from the preparation of the TPU solution and electrospinning method. The TPU was dissolved entirely in DMF at 40 °C and stirred magnetically for 12 h to get TPU solutions with different concentrations (12wt%, 14wt%, 16wt%, 18wt%, and 20wt%), named as TPU-12, TPU-14, TPU-16, TPU-18, and TPU-20.

Afterward, the TPU solution was fed into a 5 ml disposable syringe with a blunt metal needle (NO.22) clamped with a high voltage of 10 kV. The injection speed was kept at 2 mm min$^{-1}$ with the ambient temperature (40 °C) and relative humidity (25%). And the distance between the needle tip and the collector (rotating drum enveloped in Al-foil paper) was 20 cm. Finally, the TPU nanofiber membrane was dried at 70 °C under vacuum 12 h to remove the solvent. A transparent TPU film was also obtained by blade-coating the TPU solution on a clean glass substrate for comparison.

*2.3 Measurements and characterizations*

The microstructure of the TPU nanofiber membrane was characterized by scanning electron microscopy (SEM, SIGMA 300, ZEISS, Germany) at a voltage of 10 kV and microscope (DM4 P, Leica, Germany). Based on SEM images, the diameter distribution of the TPU nanofibers was characterized by Nano Measure1.2. The complex refractive index ($n$, $k$) was measured by ellipsometer (V-VASE and IR-VASE, J.A. Woollam USA), and the scattering efficiency of TPU nanofibers with different sizes in the air was established by finite-difference time-domain (FDTD) numerical simulation. The characteristic functional groups of Es-TPU membrane were characterized by ATR Fourier transform infrared spectroscopy (FT-IR, Nicolet 50, Thermo Fisher Scientific USA). The mechanical properties of the TPU nanofiber membrane were tested by a universal tensile testing machine (AGXplus, Shimadzu, Japan). The UV–vis–NIR spectral reflectance was tested by a UV–vis–NIR spectrophotometer (Lambda-1050+, Perkin Elmer, USA) in the wavelength range of 0.25-2.5 µm with an integrating sphere with a $BaSO_4$ baseline reference. The FT-IR spectrometer (VERTEX 70, Bruker, USA) with an A562 integrating sphere was used to determine the MIR spectral emittance of TPU nanofiber membranes in the wavelength range of 2.5-25 µm. The contact angle tester (JY-82A, Chengde Dingsheng experimental machine testing equipment Co., Ltd, China) was used to take the images of the water contact angle (CA) of the material. According to IOS3768-1976 and ISO 4892-3 standards, salt spray tester (SY-9, Dongguan Tianshuo Test Equipment Co., Ltd, China) and UV accelerated aging tester (UV600, SANWOOD, China) were used to carry out aging resistance test, respectively.

*2.4 Thermal Measurement*

The Solar Power Meter pyranometer (TES 1333R, TES Electrical Electronic corp., China) measured the solar radiation power during outdoor testing. Ambient humidity and wind speed can be read from the hygrometer and anemometer. A data logger thermometer with thermocouples (EX4000, Yili (Shenzhen) Technology Co., Ltd, China) was used to collect the real-time temperature measurements. A self-made device including thermocouples, a temperature signal conversion module, an aluminum foil-coated polystyrene foam box topped with PE film was used for investigating the ambient cooling performance of Es-TPU in Harbin, China (45°43′49″N, 126°38′11″E, Altitude 128 m) under direct sunlight conditions. Five identical rectangular pockets (6 cm × 8 cm × 3 cm) were prepared on top of the aluminum foil-coated polystyrene foam box for placing test samples (3 cm × 3 cm) and thermocouples. The device was placed on a shelf above the roof to reduce the heat convection from the ground.

## 3. Results and Discussion

The refractive index contrast of TPU with air and its negligible extinction coefficient (Fig. 1a) over the entire solar wavelength range endows the electrospinning membrane with a strong scattering ability with minimum absorption of sunlight [1, 50]. In the atmospheric window, the extinction peaks associated to the different vibrational modes of TPU are responsible for the strong emissivity at thermal infrared wavelengths [41, 51-53], with strong contributions from the C-C, C-O-C, and C-H bonds on the benzene ring [49, 54] (Fig. S1). The solar scattering performance of a TPU nanofiber can be studied numerically using Mie theory [1, 4, 55], revealing a high scattering efficiency for nanofiber diameters comparable to visible wavelengths (Fig. 1b), especially in the 0.3-1.0 μm wavelength range [41]. Therefore, different concentrations and spinning rates of the solution were explored to match the ideal polydispersity of the diameter distribution. Fig. 1c-d and Fig. S2 show that the average diameter and distribution width grew with increasing TPU content. This is due to the restricted motion of the polymer segments, leading to a more viscous spinning solution inhibiting fiber drafting and thinning. On the other hand, solutions with a TPU concentration < 14% cannot form nanofibers with a uniform diameter. The highest reflectance was observed for nanofibers of TPU-18% (referred to as Es-TPU in the following) which were uniformly distributed between 0.3-1.2 μm (Fig. 1d), matching the central solar spectrum wavelengths in agreement with numerical predictions.

Visually, the white Es-TPU membrane composed of randomly arranged nanofibers (Fig. 1e inset) showed a strong solar-weighted reflectivity of 95.6% (see Equation S1) with almost no transmission. In comparison, the reflectance of a free-standing transparent TPU film (Ts-TPU) was close to 10% (Fig. 1e), illustrating the importance of nano and micro-structuration of the fibers. The optical properties of TPU nanofiber membranes with different TPU concentrations were reported in Fig. S3, showing that the formation of fibers with uniform diameters favors an overall reflectance increase. However, above 18% TPU concentration, the average fiber diameter became too wide to produce effective scattering in the UV–vis region, resulting in a decreased reflectivity across the whole solar wavelength range (Fig. S3a). Based on this observation, the reflectivity was further improved by increasing the thickness of the membrane up to a value of 345 μm as a trade-off the optical performance and material cost (Fig. S4a). As mentioned above, the characteristic functional groups of TPU endow it with high infrared emittance. Given the high specific surface area of electrospinning membranes, a sufficiently thick sample can enhance the absorption of infrared radiation (Fig. S4b) compared to an unstructured TPU film owing to the reduced index mismatch at the interface of the porous medium, and the increased diffusion within the membrane itself [56]. Fig. 1f shows the emittance of Es-TPU in the thermal infrared region, reaching 93.3% in the infrared transparency window (see Equation S2). For the same reason, we observed only minimal variations in the emissivity when comparing different electrospinning membranes with comparable thickness (320±25μm) and different TPU concentrations for the spinning solution (Fig. S3b), suggesting that the thickness and specific surface area are the dominant factors determining the infrared emissivity, and that thermal infrared wavelengths show little dependence on the fine structural details due to their much larger length scales compared to the beads and fiber diameters [57]. On the other hand, the size and uniformity of the fiber diameter were also closely related to its mechanical properties. For concentrations lower than 16%, the formation of beads along the fibers causes an uneven distribution of the mechanical stress, which is detrimental for the overall resistance of the membrane. The larger and more uniform diameter distribution obtained at concentration of 18% and above showed better mechanical properties, exceeding an elongation of 200% before breaking (Fig. S5). A 6 mm wide and 0.3 mm thick strip of Es-TPU membrane was thus capable of lifting a 200 g weight without breaking (Fig. S6).

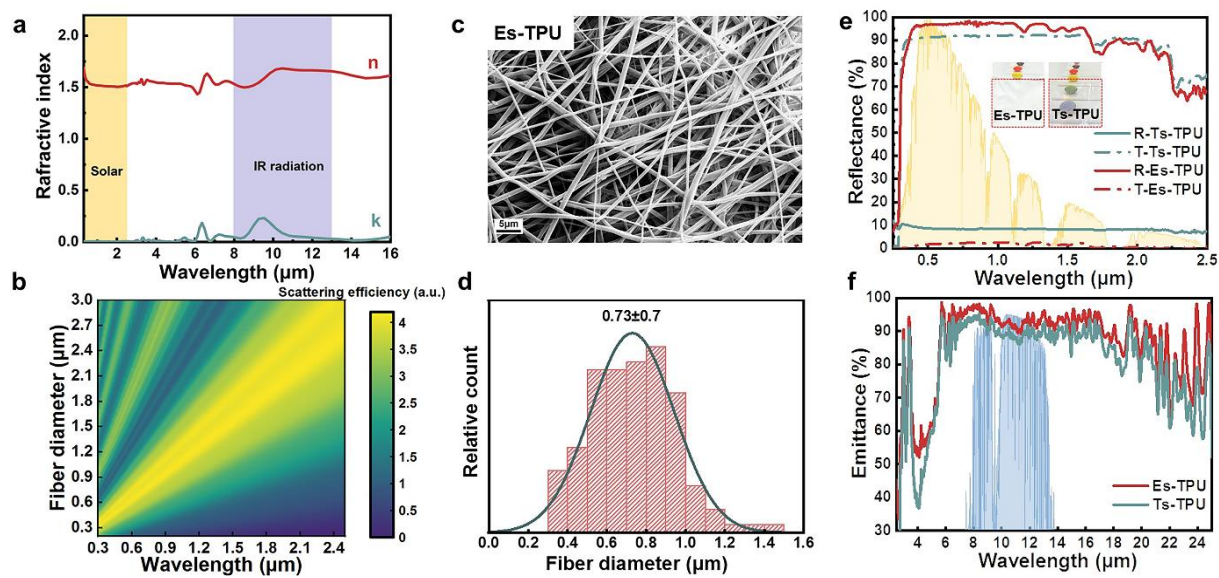

**Fig. 1.** Theoretical analysis, structures, and optical properties of Es-TPU membrane with tensile properties: a) The complex spectral refractive index ($n+ik$) of TPU in the solar and IR radiation wavelength range (0.3-16 μm), b) Simulated scattering efficiency of Es-TPU nanofibres in air in the wavelength range of 0.25-2.5 μm for different diameter values; c) Scanning electron microscopy (SEM) image of TPU-18 % nanofibres arranged to form the Es-TPU membrane;

d) Statistical distribution of the diameters of the Es-TPU nanofibre shown in Fig. 1c; e) UV–vis–NIR reflectance (solid line)/transmittance (dotted line) of Es-TPU membrane (red) and Ts-TPU film (blue) presented against the AM1.5 solar spectrum; insets: photographs of Ts-TPU and Es-TPU; f) IR emittance of Es-TPU membrane (red) and Ts-TPU film (blue) presented against the atmospheric transparency window.

Owing to the remarkable mechanical properties of the Es-TPU membrane, it is possible to adjust its optical properties by mechanical stretching. We tested the UV–vis–IR reflectance and IR emittance under strains between 0-80% (Fig. 2a and Fig. S7). The spectral reflectance and emittance measured under different degrees of strain are displayed in Fig. 2a. Fig. 2b plots the corresponding reflectivity and emissivity values. Between strain levels of 0-20%, the reflectivity decreased slowly, remaining above 91.6%. For larger strains, the reflectivity dropped more or less linearly reaching 61.1% at 80% strain, leading to a significantly increased transmittance in the solar range (Fig. S8). In contrast, the IR emissivity of the Es-TPU membrane in the atmospheric window declined more slowly, remaining around 80% over the entire range of applied strains, due to the inherent high infrared absorption of TPU. Fig. 2c and Video S1 further show the effect of increasing the transparency of the Es-TPU membrane after mechanical stretching. In its initial state the membrane had a white appearance (Es-TPU-0%) with almost no transmittance, while at an 80% strain (Es-TPU-80%), the underlying logo became partially visible due to the increased transmissivity of the membrane. Stable reflectivity recovery was observed after 100 actuation cycles from 80% strain to 0% (Fig. S9). Notably, the optical properties of the membrane are quickly restored when the tensile force is removed.

A practical application scenario is presented in Fig. 2d. When the Es-TPU membrane is used in the outdoor parking lot or building roof, internal objects will absorb the transmitted sunlight, and the luminous flux can be adjusted according to the requirements of personal comfort through the mechanical deformation of the membrane. Moreover, the random internal structure of the nanofiber membrane exhibited negligible angular dependence of reflectivity at all strain levels (Fig. S10), thus avoiding glare and increasing the utility of the membrane [58].

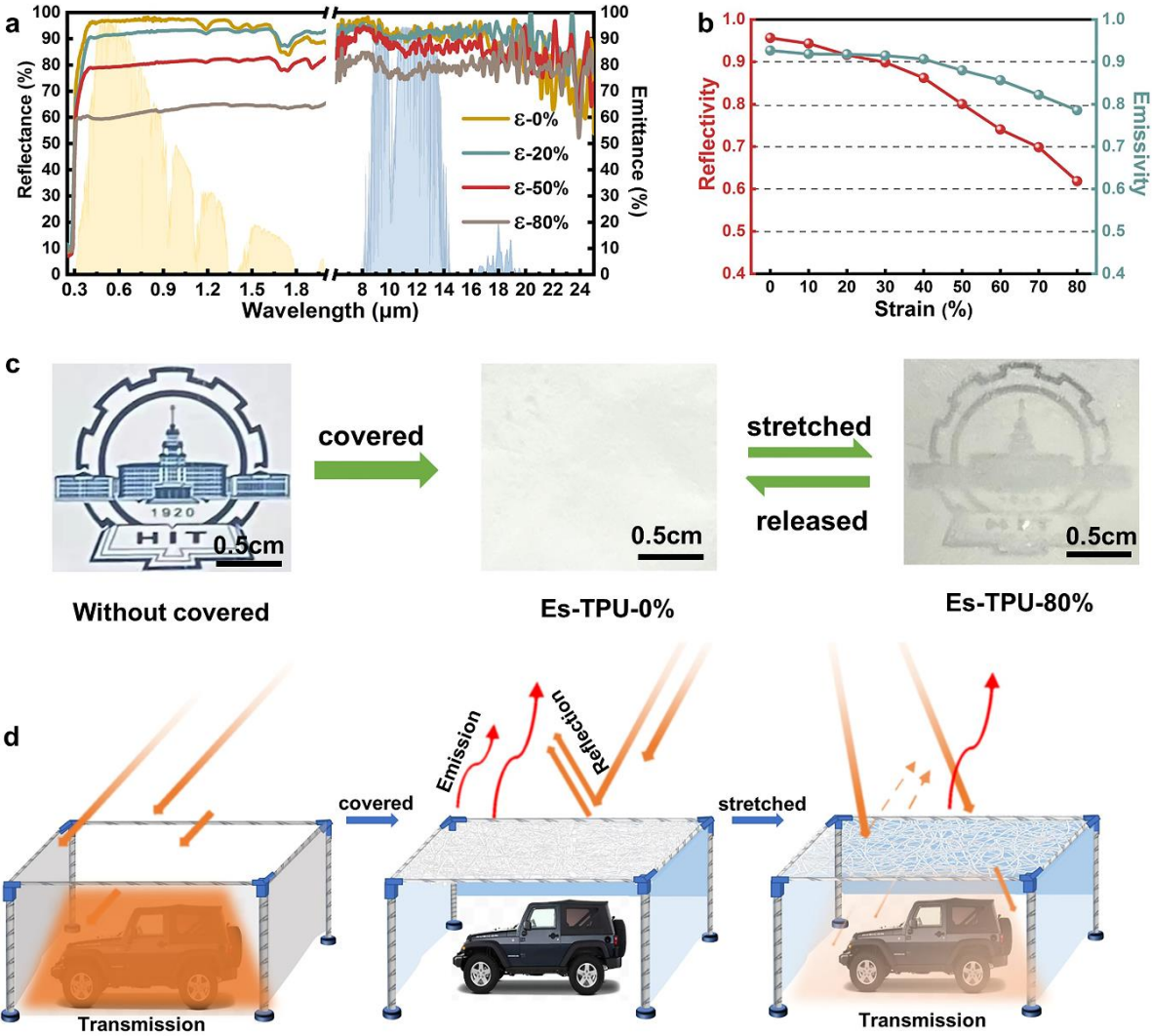

**Fig. 2.** The optical properties of Es-TPU membrane under different strains: a) UV–vis–IR reflectance/emittance of Es-TPU membrane with different strains (0%, 20%, 50%, and 80%) presented against the AM1.5 solar spectra (yellow) and the atmospheric transparency window (blue); b) The variation trend of reflectance and emittance of Es-TPU membrane with different strains; c) Photographs of the substrates covered by nothing (left), Es-TPU-0% (middle), and Es-TPU-80% (right); d) Schematic illustration of the optical regulation of Es-TPU membrane.

To reveal the reasons for the changes in optical property of the Es-TPU membrane in Fig. 2a, the microstructure of the membranes under different strains was characterized by SEM (Fig. 3a-3d and Fig. S11). While the observed nanofiber diameter distribution remained largely unchanged under deformation, large, micron-size pores (measured as the average distance between parallel fiber traits) appeared even at low strain levels. The strain applied on the membrane caused the nanofibers to straighten along the deformation direction and to bend along the perpendicular direction (Fig. 3b). At the same time, the deformation was associated to a decrease of the membrane thickness and an increased pore size leading to a lower area density of nanofibers which reduced the available scattering interfaces, and thus to a larger transmittance. Above strain levels of 50%, macro-pores with sizes approaching those of thermal infrared wavelengths started to appear, which also explained the gradual decrease in thermal emissivity observed above this degree of elongation as the radiation in this wavelength range became more transmitted and had a lower chance of getting absorbed via internal diffuse reflectance [59]. However, the overall impact of these effects on IR emissivity is attenuated by the intrinsic high absorption of TPU in the infrared range, leading to a less pronounced change compared to that in the solar range.

Based on SEM measurements, the two main mechanisms responsible for the increased transmittance upon stretching were identified in the rearrangement of the nanofibers into a network with reduced density (see sketch in Fig. 3e) and the overall reduction of the membrane thickness, which we measured in Fig. 3f. The scattering efficiency ($Q_{sca}(\lambda)$) in Equation 1 further proves that the scattering efficiency is inversely proportional to the size of the empty regions between the nanofibers, and will be significantly reduced when the pores size increases [60]. For a finite-thickness porous polymer membrane with low absorption and isotropic scattering, the spectral reflectance ($R_{sca}(\lambda)$) can be approximated as (Equation 2) [61-63]. In our case, a compound effect is obtained considering that the membrane elongation has a combined effect of both decreasing the sample thickness, and make the nanofiber network sparser, hence weakening its optical scattering strength (Equation 3).

$$Q_{\text{sca}}(\lambda) = \frac{\sigma_{\text{sca}}(\lambda)}{\frac{1}{4}\pi d^2} \qquad (1)$$

where *d* is the diameter of ascattering nanopore, and $\sigma_{sca}(\lambda)$ is its scattering cross-section. The approximated reflectance can be thus written as

$$R_{\text{sca}}(\lambda) = \frac{\frac{3}{4}\rho\sigma_{\text{sca}}(\lambda)t}{1+\frac{3}{4}\rho\sigma_{\text{sca}}(\lambda)t} \qquad (2)$$

where *λ* is the wavelength in vacuum, *ρ* is the number density of nanopores, and *t* is the thickness of the membrane, so that Equation 3 can be finally obtained

$$R_{\text{sca}}(\lambda) = 1 - \frac{1}{1+\frac{3}{4}\rho\sigma_{\text{sca}}(\lambda)t} \qquad (3)$$

For our thermal management application, these two effects further add up to the fact that less thick media are also associated with a lower thermal emissivity (see Fig. S4b), leading to an even more effective solar heating effect.

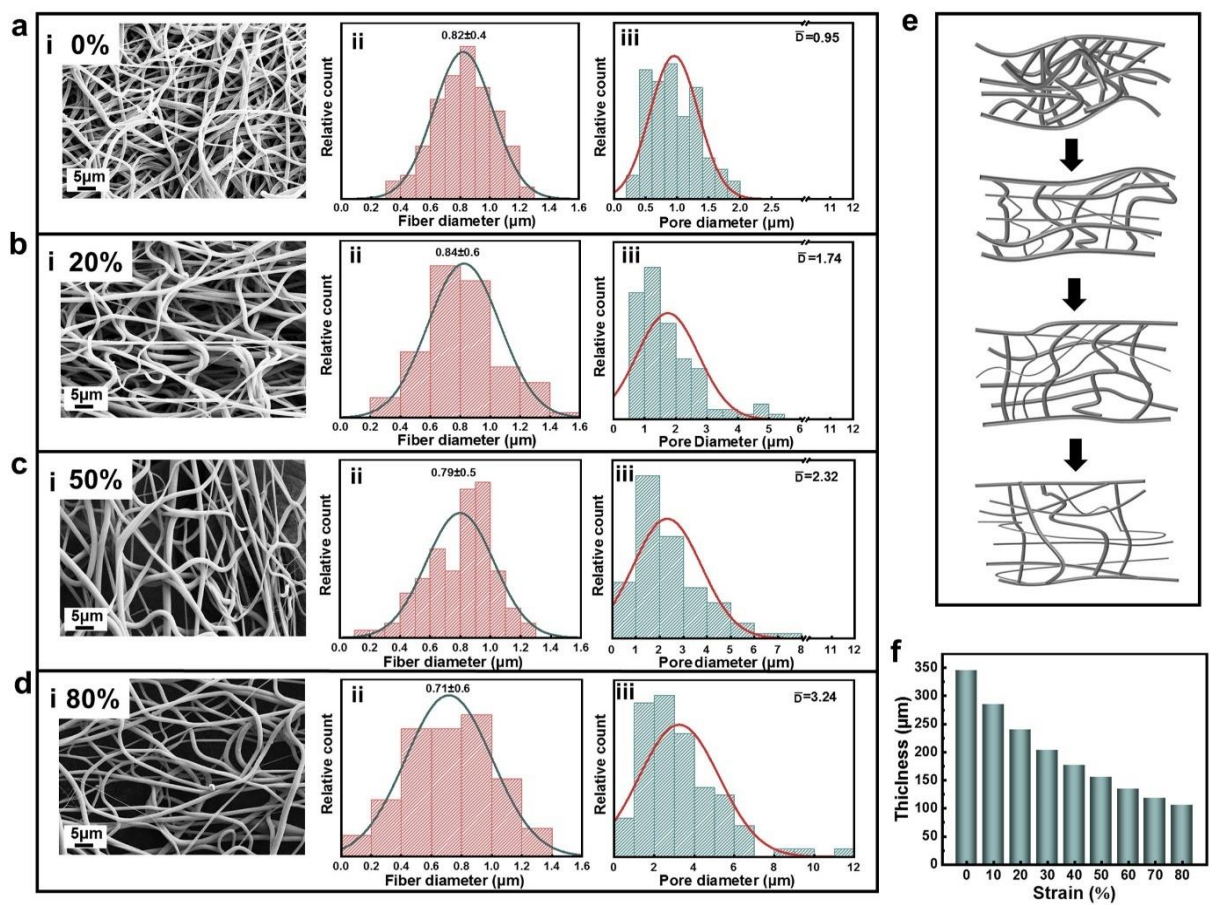

**Fig. 3.** Morphological changes of Es-TPU membrane under different strains: Panels a), b), c) and d) refer to membranes under 0%, 20%, 50% and 80% strain levels, respectively, i: scanning electron microscopy (SEM) images, ii: corresponding nanofiber and iii: pore size distributions of Es-TPU membrane; e) Schematic illustration of the morphological changes in Es-TPU membrane during tensile strain; f) Thickness variation of Es-TPU membrane with different strains.

The temperature control capability of the Es-TPU membrane under different strains was demonstrated in an outdoor cooling test (Fig. 4a) on the rooftop of the eighth floor under direct sunlight (Harbin Institute of Technology, 45°43′49″N, 126°38′11″E), with weather conditions reported in Fig. 4b for the measurement day (July 22$^{nd}$ 2022). A polystyrene foam sample box wrapped with aluminum foil and polyethylene film was used to shield the measurement system from thermal radiation and convection (Fig. 4c). The real-time temperature and the temperature difference ($\Delta T$) with the ambient environment and the black aluminum sheets (high absorption) placed under the Es-TPU membranes at 0%, 20%, 50%, and 80% strains are shown in Fig. 4d and Fig. 4e, respectively. The Es-TPU-0% exhibited good cooling performance with a 10 °C drop in temperature under a solar intensity ($I_{solar}$) of 900 W m$^{-2}$ at 12:00-13:00. After 15:00, the decreased irradiance accompanied by the increased relative humidity resulted in a weakened cooling capacity [64]. The cooling performance of the Es-TPU membrane is attributed to its high solar reflectivity (reducing heat gains) and high infrared emissivity (increasing radiative losses). As the degree of strain increased, the cooling capacity of Es-TPU membrane gradually decreased, and the solar transmitted light is eventually converted into heat on the substrate. At a strain level of 50% (Es-TPU-50%), the temperature of the substrate covered by the membrane was almost the same as the ambient temperature, that is, the cooling capacity of the Es-TPU-50% was offset by the transmitted sunlight. With even higher strain levels, a maximum temperature increases of 9.5 °C was finally observed for Es-TPU-80% compared to the ambient environment (Fig. 4e).

The theoretical cooling power attainable by the Es-TPU membrane was calculated using equations reported in Supplementary Note 2. The resulting $P_{cool}$ of Es-TPU-0 % at different nonradiative heat transfer coefficients ($h_{cc}$) relative to conduction and convection under direct sunlight is displayed in Fig. 4f. When the temperature of the Es-TPU membrane equaled ambient temperature ($T_{amb}$ = 303 K), the cooling power was 118.25 W m$^{-2}$. With increasing $h_{cc}$ (0-3-6-9-12 W m$^{-2}$ K$^{-1}$), the obtained maximum $\Delta T$ were progressively reduced to 32.2 °C, 17.1 °C, 11.7 °C, 8.8 °C and 7.3 °C, respectively. We also calculated the $P_{cool}$ of Es-TPU-0% at different ambient temperatures ($T_{amb}$), which showed how $P_{cool}$ increases at higher $T_{amb}$ following the blackbody radiation law (Fig. 4g) [65]. The $P_{cool}$ of the Es-TPU membranes at $h_{cc}$ = 12 W m$^{-2}$K$^{-1}$ [66] (Fig. 4h and Fig. S12) and the max $\Delta T$, radiative cooling and solar heating powers (Table S1) under different strains were also calculated. The cooling performance of the membrane vanished approaching a strain level of 50%, beyond which it entered the solar heating stage reaching eventually a maximum power of 220.34 W m$^{-2}$ at 80% strain. The largest rate of variation for the reflectivity is observed in the strain range between 0% and 80%, with progressively smaller improvement for more extreme deformations. Considering that the solar heating measured at 80% strain (well below the 200% maximum elongation at break) is already significant, we select this value as the best tradeoff between heating power and service life / cyclability of the material. These results demonstrated that the Es-TPU membrane offered a broad daytime temperature tuning range (between -10 °C and +9.5 °C), which can be adjusted taking advantage of its good tensile properties according to the changes in ambient temperature to maintain thermal comfort.

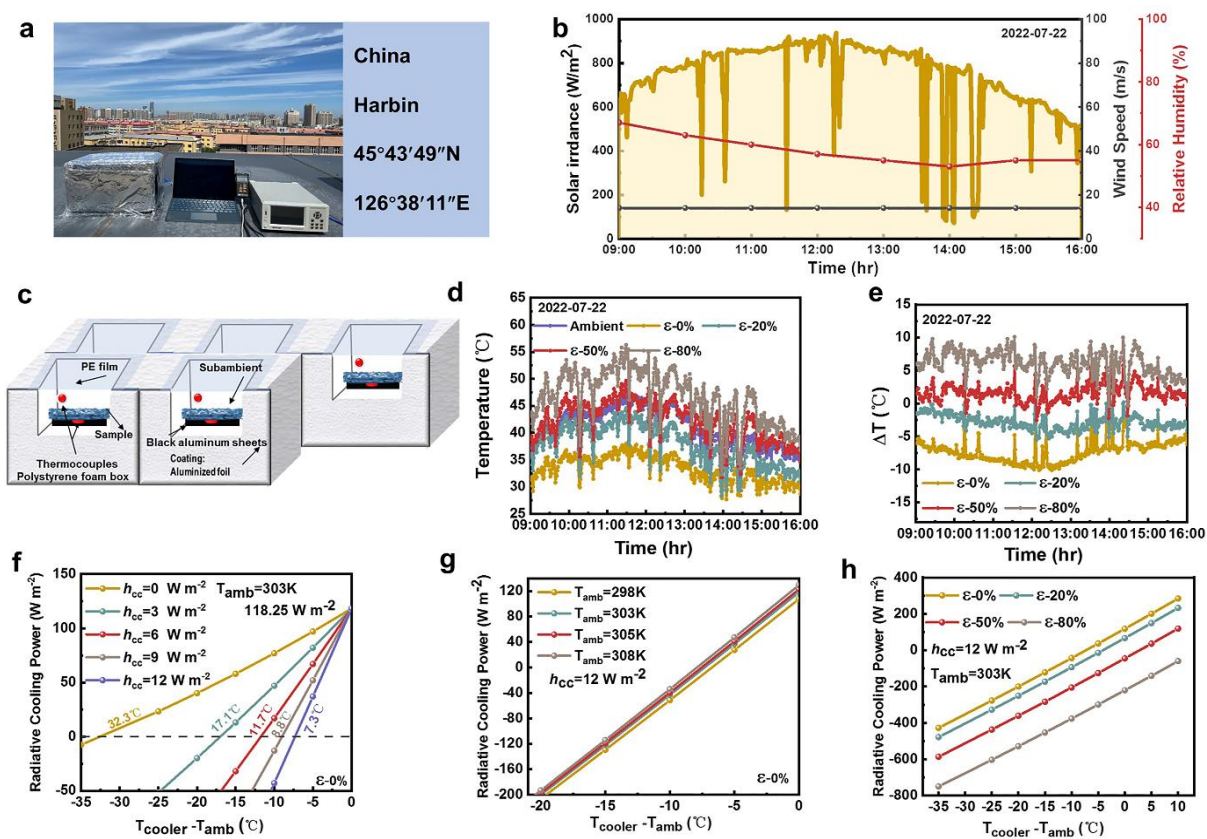

**Fig. 4.** Thermal measurement and radiative cooling performance of Es-TPU membrane under different strains: a) Photograph of the outdoor test environment (Harbin, 45°43′49″N, 126°38′11″E, China); b) Solar irradiance, relative humidity, and wind speed date of the test location (July 22nd 2022, 09:00-16:00); c) Schematic illustration of the assessment device; d)

Real-time temperature curves of the outdoor experiment with the Es-TPU membrane under different strain levels; e) The temperature difference ($\Delta T=T_{cooler} - T_{amb}$) curves of the data in d); f) Theoretical radiative cooling power of Es-TPU membrane under different $h_{cc}$ (0-12 W m$^{-2}$) ($T_{amb}$=303 K); g) Theoretical radiative cooling power of Es-TPU membrane at different ambient temperature ($h_{cc}$=12 W m$^{-2}$).; h) Theoretical radiative cooling/ solar heating power of Es-TPU membrane under different strains (0 %, 20 %, 50 %, 80 %) ($h_{cc}$=12 W m$^{-2}$, $T_{amb}$=303 K).

Due to its targeted outdoor application, the resistance of the Es-TPU membrane should also be tested against external agents such as rain, erosion, sunburn, and dust deposition, which may degrade its optical performance. We investigated the influence of the extreme outdoor environment conditions on the reflection / emission capability of the membranes. Anti-adhesion performance tests showed that the membrane is in a hydrophobic state (the contact angle is 122°), such that any liquid droplets could be easily removed from the sample surface leaving no residue, which is likely due to the low surface energy of the hierarchically rough structures (Fig. 5a). Consequently, the Es-TPU membrane can be easily rinsed with water to remove accumulated dust or soil. (Fig. 5b, Video S2). Also, other droplets of liquids with different surface tensions were shown to maintain a spherical shape, revealing excellent liquid resistance (Fig. 5c). The reflectivity of the Es-TPU membrane with soil attached to the surface decreased significantly, especially in the UV–vis due to the absorption of sunlight by the soil, but is restored upon cleaning (Fig. 5d). In contrast, soil contamination affected negligibly the infrared emissivity (Fig. S13a). Notably, the Es-TPU membrane exhibited good stress and strain recovery after 10,000 tensile cycles between 0% and 80% unidirectional strains (Fig. S14), with a small reflectivity drop from 95.6% to 92.5% (Fig. 5e) and no significant change in the IR emittance (Fig. 5f). Optical microscopy in reflectance mode (Fig. 5g) revealed the formation of micro-cracks (marked by a red dotted line) perpendicular to the tensile direction on the surface of the Es-TPU membrane after the tensile cycles, which was due to the mismatch in mechanical properties between the different nanofibers that constitute the membrane [49, 67, 68].

In addition to its deformability, the membrane is also expected to maintain its properties under harsh conditions for practical applications. The chemical durability and UV stability of the hydrophobic Es-TPU membranes were evaluated by exposing it to a neutral salt spray environment, UV irradiation, and simulated acid rain (pH=2). The reflectance at 0.25-2.5 μm wavelength reduced from 95.6% to 92.9% after 24 h of neutral salt spray treatment, retaining a good reflectivity (Fig. 5h). Prolonged UV exposure caused a gradual increase of TPU absorption at shorter wavelengths, with a consequent drift of the membrane appearance towards a slightly yellow tone (Fig. 5i inset) due to the intrinsic UV absorption of TPU. Even in this case, the reflectivity of the Es-TPU membrane in the visible range remained above 90 % (Fig. 5i). At an irradiation power of 400 W h$^{-1}$ in Heilongjiang, a 72 h UV exposure was equivalent to 36 days of continuous outdoor exposure (see Supplementary Note 3). However, the cost per unit area of the Es-TPU membrane (15 USD m$^{-2}$, see Supplementary Note 4) could allow to replace it whenever the optical performance decreased below a minimum threshold. The Es-TPU membrane exhibited also good resistance to simulated acid rain (Fig. S15), with just a partial degradation in the IR range (Fig. S13b and c), due to the degradation of the chemical bonds of TPU. Finally, we tested the mechanical durability of the Es-TPU membrane by sandpaper abrasion. A weight of 500 g was placed on the membrane and moved cyclically for 100 cm on a 120-grit sandpaper, with negligible effect on its hydrophobicity (Fig. S16 a-b) and overall spectral properties (Fig. S16 c and d). The haze test results showed that the Es-TPU-80% allowed even natural illumination of an internal space, with a high value of 93 % (Fig. 5j) ensuring both uniform lighting and privacy (Fig. S17) [69, 70]. The good durability exhibited by Es-TPU membranes under different conditions confirmed its potential in practical application as a thermal control material.

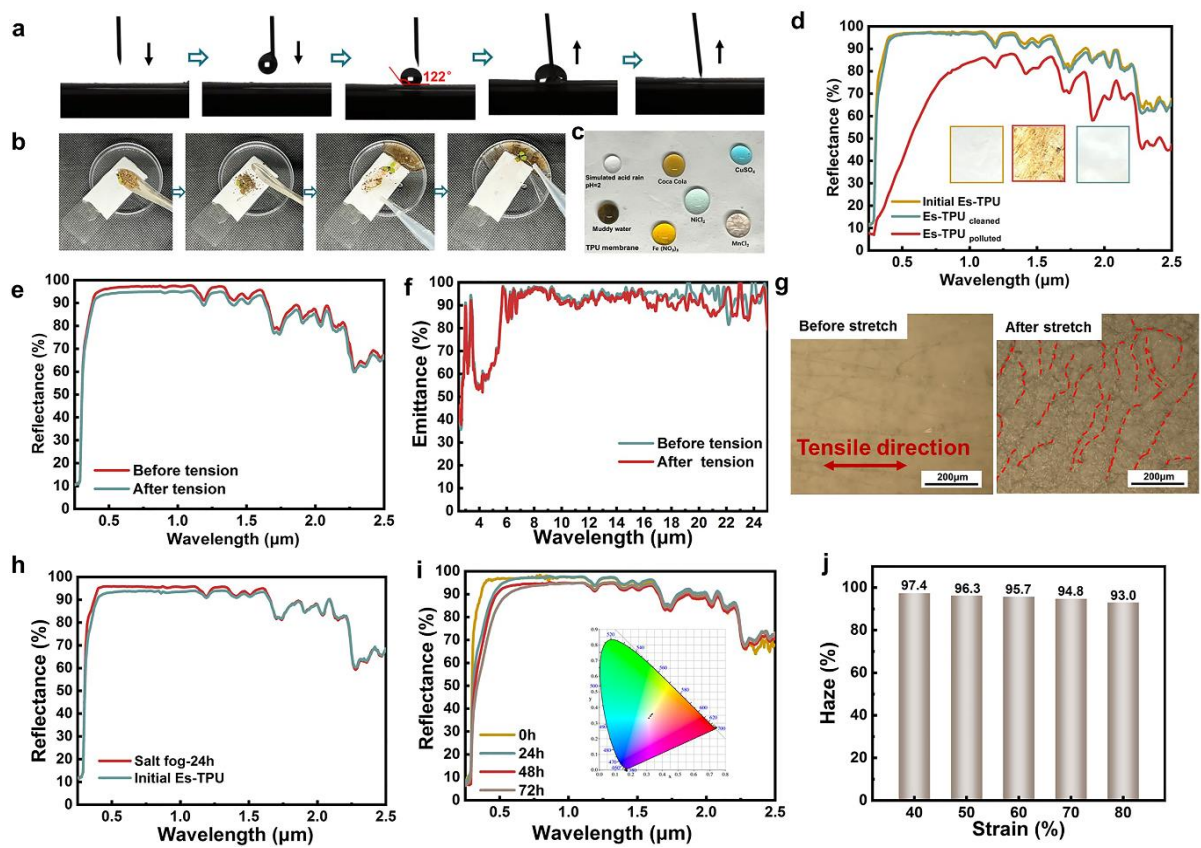

**Fig. 5.** Weatherability of the Es-TPU membrane: a) Hydrophobicity and b) Photographs of a droplet flowing over the Es-TPU membrane and taking away the soil; c) Anti-adhesion tests with different liquid agents. UV–vis–NIR reflectance of Es-TPU membranes d) initial, soiled and cleaned, insets: optical images of the three states; e) UV–vis–NIR reflectance of Es-TPU membrane before (blue) and after (red) 10,000 times strain of 80%; f) IR emittance of Es-TPU membrane before (blue) and after (red) 10,000 times strain of 80%; g) Optical images of Es-TPU membrane before and after 10,000 strain cycles to 80%. h) before and after salt spray treatment for 24 h; i) before and after 0 h, 24 h, 48 h, 72 h of UV aging test, inset: CIE 1931 chromaticity diagrams; j) Haze of Es-TPU membrane under different strains.

## 4. Conclusions

In summary, we prepared an in-situ solvent-free TPU nanofiber membrane with adjustable optical properties (ΔR=34.5%) by electrospinning. The membrane exhibited 95.6% reflectivity (0.25-2.5 μm) and 93.3% infrared emissivity (8-13 μm) in its rest configuration (radiative cooling mode), reaching a maximum theoretical cooling power of 118.25 W m$^{-2}$. Varying the degree of mechanical stretching, continuous adjustment between radiative cooling and solar heating could be achieved ranging from a -10 °C temperature drop for 0% stretch to +9.5 °C at an 80% elongation. Such mechanical stretching can be achieved either manually of with a motorized rotor, similarly to what is already done with common window shutters and awnings (see Supplementary Note 5 for an estimation of the actuation energy needs). Once in their desired configuration, the membranes can exert their thermal management action without further energy inputs, offering a a much needed continuously tunable solution meeting the practical demand for all-weather outdoor thermal management.

**Acknowledgements**

The work has been supported by the Fundamental Research Funds for the Central Universities (Grant No. HIT.OCEF.2021004, FRFCU5710090220), the National Natural Science Foundation of China (No. 51702068, 52072096), and the Heilongjiang Postdoctoral Fund (LBH-Z15078, LBH- Z16080). Part of this work is supported by the European project PaRaMetriC, code 21GRD03. The project 21GRD03 PaRaMetriC received funding from the European Partnership on Metrology, co-financed by the European Union's Horizon Europe Research and Innovation Programme and from the Participating States. G.E.L. acknowledges support from the "FSE-REACT EU" program financed by National Social Fund–National Operative Research Program and Innovation 2014-2020 (D.M. 1062/2021), personal Grant number 10-G-15049-2.

**Declaration of interests**

The authors declare no competing interests.

**Supporting Information**

**Data and code availability**

Data will be made available on request.

*Supporting Information for*

# Strain-adjustable Reflectivity of Polyurethane Nanofiber Membrane for Thermal Management Applications

Xin Li[a], Zhenmin Ding[a], Giuseppe Emanuele Lio[b,c], Jiupeng Zhao[a], Hongbo Xu[a, *], Lorenzo Pattelli [b,d *], Lei Pan[a, *], Yao Li[e, *]

[a] School of Chemistry and Chemical Engineering, Harbin Institute of Technology, Harbin, 150001, PR China

[b] European Laboratory for Non-Linear Spectroscopy (LENS), Sesto Fiorentino, 50019, Italy

[c] Department of Physics and Astronomy, University of Florence, Sesto Fiorentino, 50019, Italy

[d] Istituto Nazionale di Ricerca Metrologica (INRiM), Turin, 10135, Italy

[e] Center for Composite Materials and Structure, Harbin Institute of Technology, Harbin, 150001, China

*Corresponding author: iamxhb@hit.edu.cn, l.pattelli@inrim.it, panlei@hit.edu.cn, liyao@hit.edu.cn

**Supplementary Note 1.** Definition of average solar reflectance and infrared emittance.

The average solar reflectance in the full solar spectrum (0.25-2.5 μm) can be calculated by equation S1:

$$\overline{R}_{solar} = \frac{\int_{0.25}^{2.5} I_{solar}(\lambda) R(\lambda) d\lambda}{\int_{0.25}^{2.5} I_{solar}(\lambda) d\lambda} \tag{1}$$

where $I_{solar}(\lambda)$ is the solar intensity spectrum at air mass (AM) 1.5, and $R(\lambda)$ is the spectral reflectance in the in the full solar spectrum.

The average IR emittance in the atmospheric window (8-13 μm) can be calculated by equation S2:

$$\varepsilon_{8\text{-}13\mu m} = \frac{\int_{8}^{13} I_{BB}(T,\lambda) \varepsilon(T,\lambda) d\lambda}{\int_{8}^{13} I_{BB}(T,\lambda) d\lambda} \tag{2}$$

where $I_{BB}(T, \lambda)$ is the spectral intensity emitted by a standard blackbody with temperature, and $\varepsilon(T, \lambda)$ is the spectral emittance of a cooler.

According to Kirchhoff's law [1], $\varepsilon(T, \lambda)$ can be defined by equation S3:

$$\varepsilon = 1\text{-}T\text{-}R \tag{3}$$

where $T$ is the ransmittance and $R$ is the reflectance in IR spectroscopy.

**Supplementary Note 2.** Theoretical calculation of daytime radiative cooling power

When the radiation cooling device is placed under the sun, the net cooling power $P_{net}$ depends on the following equation:

$$P_{net}(T) = P_{rad}(T) - P_{atm}(T_{amb}) - P_{sol} - P_{(cond+conv)} \quad (4)$$

Where, $P_{rad}$ is the thermal outward radiative power by the radiative cooler. $P_{atm}$ is the absorbed energy from atmospheric thermal radiation, $P_{sol}$ is the absorbed incident solar radiation, and the $P_{cond+conv}$ is the heat conduction-convection from ambient radiation.

**Thermal outward radiative power ($P_{rad}$):**

$$P_{rad}(T) = \int d\Omega \cos\theta \int_0^\infty d\lambda I_{BB}(T,\lambda)\varepsilon(\lambda,\theta) \quad (5)$$

Where, $\theta$ is the local zenith angle, $I_{BB}(T,\lambda)$ is the intensity of the radiation wave at the real-time temperature and the wavelength generated by the blackbody. $\varepsilon(\lambda,\theta)$ is the emissivity of the material at the wavelength $\lambda$.

angular integral on the hemisphere:

$$I_{BB}(T,\lambda) = \frac{2hc^2}{\lambda^5} \frac{1}{e^{\frac{hc}{\lambda T k_B}} - 1} \quad (6)$$

Where $T$ is spectral radiation temperature of blackbody, $h$ is the Planck's constant. $k_B$ is the Boltzmann constant, $c$ is the speed of light, and $\lambda$ is the wavelength.

**Absorbed energy from atmospheric thermal radiation ($P_{atm}$):**

$$P_{atm}(T_{amb}) = \int d\Omega \cos\theta \int_0^\infty d\lambda I_{BB}(T_{amb},\lambda)\varepsilon(\lambda,\theta)\varepsilon_{atm}(\lambda,\theta) \tag{7}$$

Where $T_{amb}$ is the ambient temperature, $\varepsilon_{atm}(\lambda,\theta)$ is the atmospheric emissivity at zenith angle $\theta$ and wavelength $\lambda$.

**Absorbed incident solar radiation ($P_{sol}$):**

$$P_{sol} = \int_0^\infty d\lambda \in (\lambda,\theta_{sol}) I_{AM1.5}(\lambda) \tag{8}$$

Where $I_{AM1.5}(\lambda)$ is the AM1.5 received standard solar spectral irradiation intensity and $\theta_{sol}$ is the inclination of the sample surface facing the sun.

**Heat conduction-convection from ambient radiation ($P_{cond+conv}$):**

$$P_{cond+conv}(T, T_{amb}) = h_{cc}(T_{amb} - T) \tag{9}$$

Where $h_{cc}$ is the non radiative heat transfer coefficient which, in low humidity climate, has a typical values comprised in the range 0-12 W m$^{-2}$ K$^{-1}$ [2, 3]. The cooling power is greatly affected by the atmospheric transmittance, humidity and wind speed, heat quality, solar irradiance, etc. [4].

**Supplementary Note 3.** UV aging time conversion estimation

Here, we calculate the conversion between the test time of ultraviolet aging test chamber and the outdoor illumination time according to the indicators provided by the equipment company [5]. Outdoor lighting conditions are not only related to seasons and regions. We estimated the time equivalent to outdoor irradiation for 1 day of UV testing according to the following equations:

$$\frac{Q_{year}}{365 \times 24} = Q_{hour} \tag{10}$$

$$\frac{T_{set}}{T_{amb}} \times \frac{Q_a}{Q_{hour}} \times 1.5 = D_{outdoor} \tag{11}$$

Where $Q_{year}$ is the annual total solar radiation. $Q_{hour}$ is the annual average hourly irradiance of solar energy. $Q_a$ is the total radiation intensity set during the test. $T_{set}$ is the temperature set during the test. $T_{amb}$ is the local annual average temperature. $D_{outdoor}$ is the outdoor irradiation time.

In our test, $Q_{year}$ was 1200 kW h$^{-1}$ m$^{-2}$ in Heilongjiang Province, $T_{amb}$ was 11 °C. $Q_a$ was set as 1000 ×0.5 W m$^{-2}$, $T_{set}$ was set as 24 °C. After calculation, the $D_{outdoor}$ is 12 days. Therefore, the 3-day UV irradiation test in the experiment is equivalent to 36 days of outdoor irradiation.

**Supplementary Note 4.** Fabricated cost of the Es-TPU membrane

Price per unit:

Thermoplastic Urethane (TPU, 8795A) =35 CNY kg$^{-1}$= 0.0049 USD g$^{-1}$

N, N-dimethylformamide (DMF) = 44 CNY 500mL$^{-1}$ = 0.0123 USD mL$^{-1}$

Dosage and Price:

A 100 cm$^{-2}$ Es-TPU membrane requires 2.4g TPU and 11mL DMF

Total : 0.0049*2.4+0.0123*11=0.15 USD

Unit cost : 15 USD m$^{-2}$

**Supplementary Note 5.** Illustrative evaluation calculation of the membrane cost and benefit:

The external force required to achieve a certain strain is:

$$F = \sigma * b * d \tag{12}$$

Where $F$ is the maximum tensile load when the specimen breaks, $\sigma$ is the tensile stress, and $b$, $d$ are the width and thickness of the specimen, respectively.

It can be seen from Fig. S6 that the maximum external force required to generate 80% strain of the membrane is: 3.37 MPa*0.345 mm*3.2m=3720 N

$$W = F * L \tag{13}$$

Where $W$ is the work done by external force, $F$ is the maximum tensile load when the specimen breaks and $L$ is the moved distance.

The power required by the external force to generate 80% strain along one direction of the film is: 3720N*2.56m =9523.2 J. Considering a power consumption of 735W of an air conditioning unit running for one hour, we get an energy consumption more than 3 orders of magnitude larger, at $735*3600=2.65*10^6$ J.

**Supplemental Video 1:** Fast reflectivity switching of Es-TPU membrane through mechanical stretching.

**Supplemental Video 2:** Self cleaning ability of Es-TPU membrane: running water can wash away the soil pollutants on the surface of the membrane.

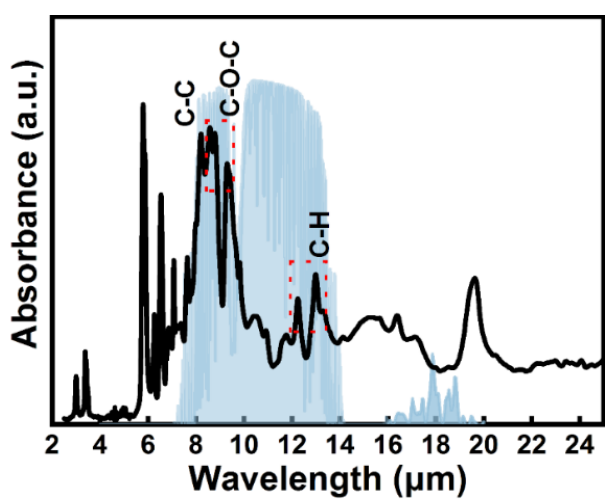

**Fig. S1.** Absorbance spectrum of Es-TPU membrane measured with ATR-FTIR spectroscopy.

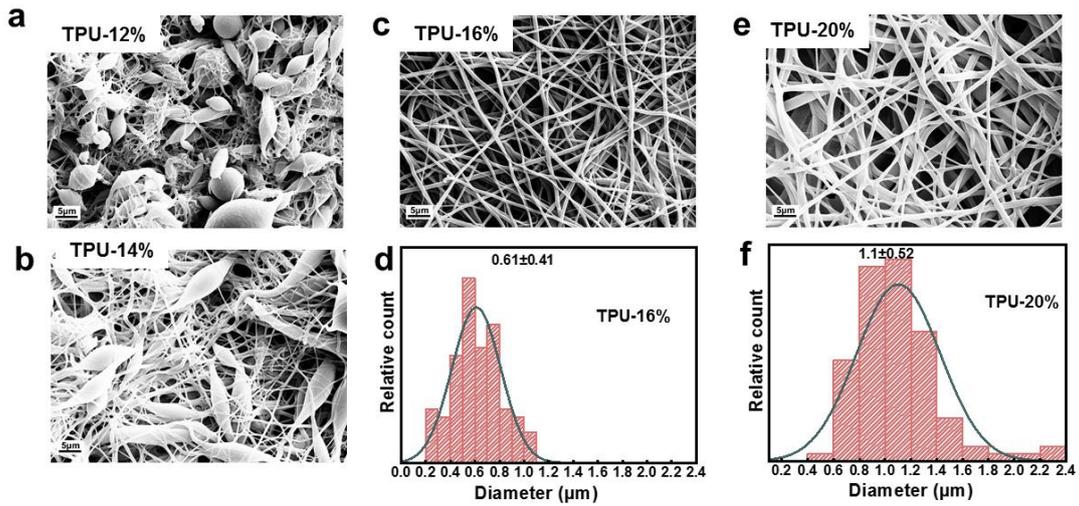

**Fig. S2.** SEM images and the diameters distribution of the Es-TPU nanofibers with different TPU concentrations: a) TPU-12%; b) TPU-14%; c) TPU-16%; d) The diameter distribution of the nanofibers in Fig. S2c; e) TPU-20%; f) The diameter distribution of the nanofibers in Fig. S2e.

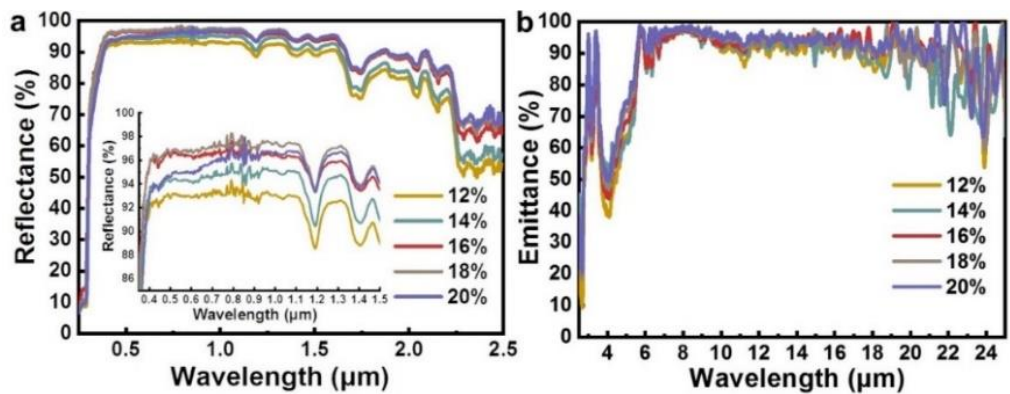

**Fig. S3.** Influence of TPU concentrations on the optical performance: a) UV–vis–NIR reflectance b) IR emittance of Es-TPU membranes with different TPU concentrations; Inset: UV–vis–NIR reflectance of Es-TPU membranes in the 0.3-1.5μm wavelengths.

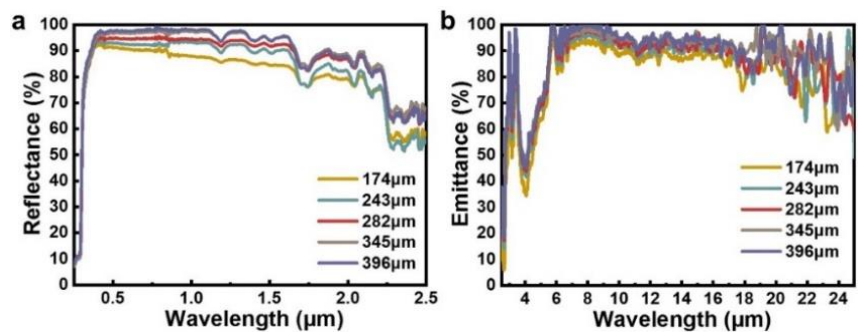

**Fig. S4.** Influence of TPU concentrations on the optical performance: a) UV–vis–NIR reflectance b) IR emittance of Es-TPU membranes with different TPU concentrations.

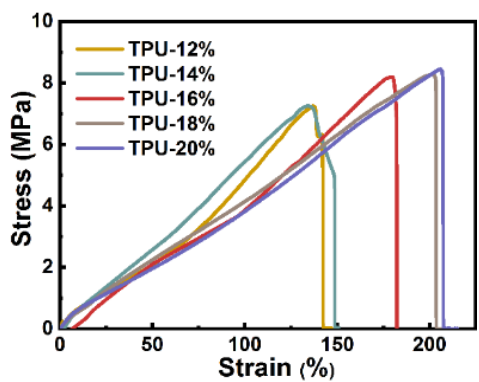

**Fig. S5.** Stress-strain curves of Es-TPU membranes with different TPU concentrations.

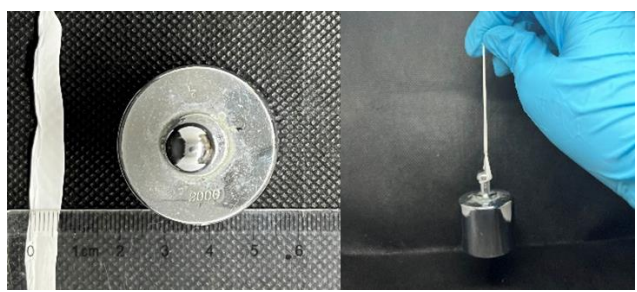

**Fig. S6.** Mechanical tensile load-bearing capacity of Es-TPU membrane.

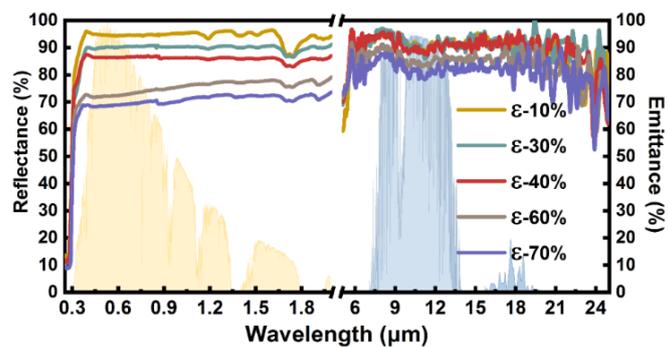

**Fig. S7.** UV–vis–IR reflectance/emittance of Es-TPU membrane with different strains (10%, 30%, 40%, 60%, 70%).

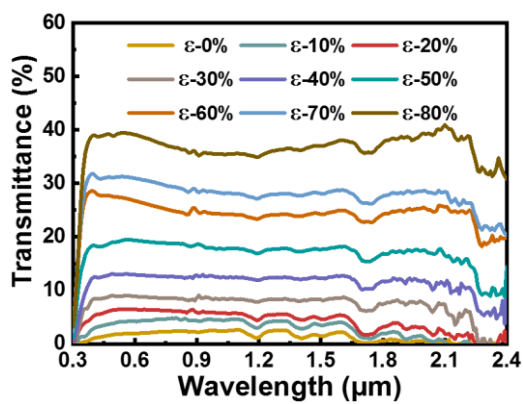

**Fig. S8.** UV–vis–NIR transmittance of Es-TPU membrane with different strains (0%-80%).

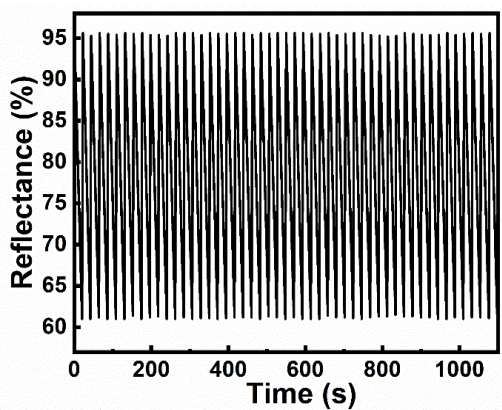

**Fig. S9.** Reflectance change as a function of stretching/release cycles.

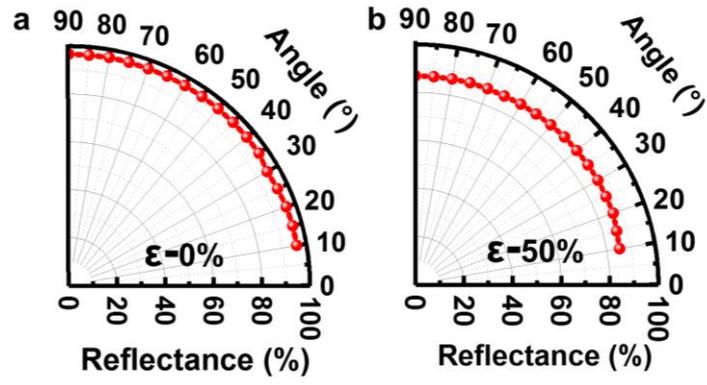

**Fig. S10.** The reflectance of Es-TPU membrane during at 550 nm wavelength at different incident angles: (a) The strain of Es-TPU membrane is 0%, (b) The strain of Es-TPU membrane is 50%.

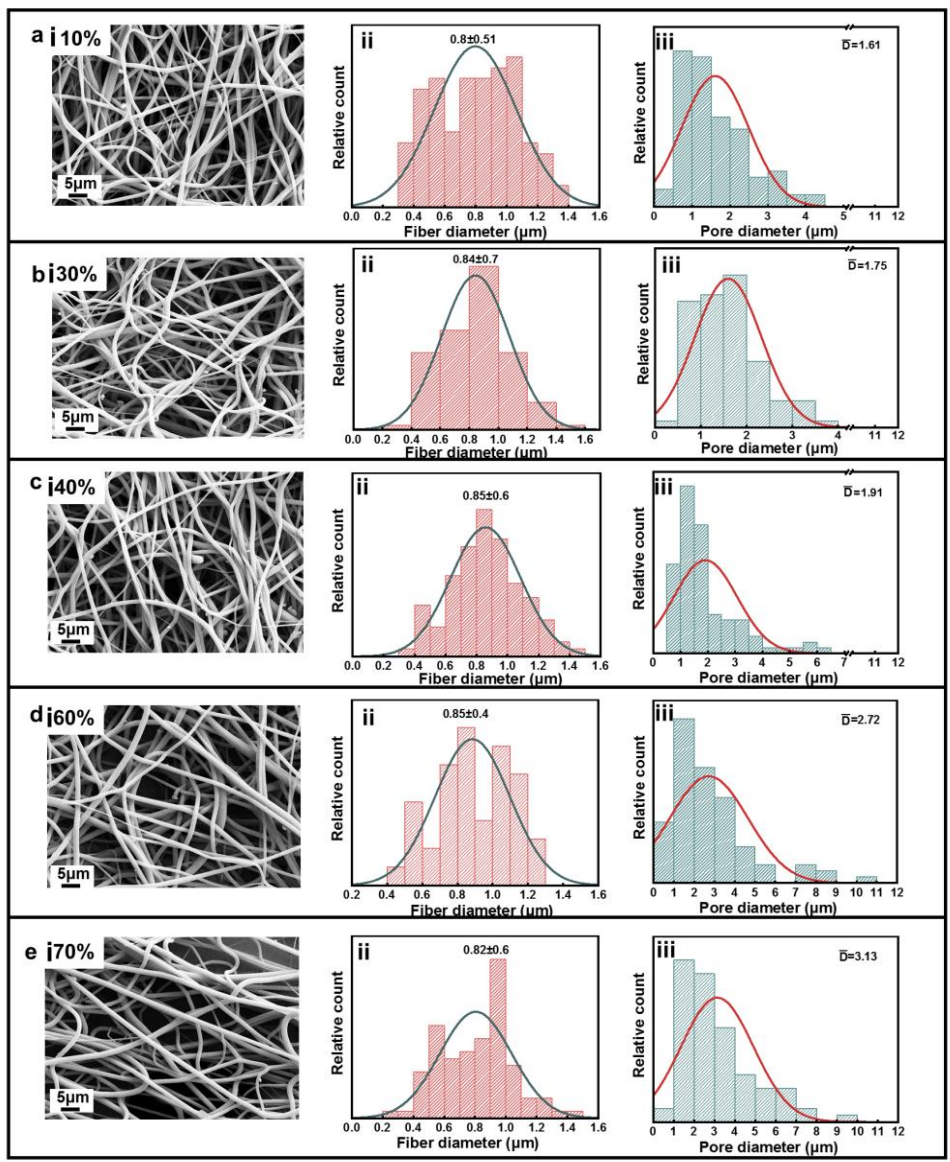

**Fig. S11.** Morphological changes of Es-TPU membrane under different strains: Panels a), b), c), d) and e) refer to membranes under 10%, 30%, 40%, 60% and 70% strain levels, respectively, i: scanning electron microscopy (SEM) images, ii: corresponding nanofiber and iii: pore size distributions of Es-TPU membrane.

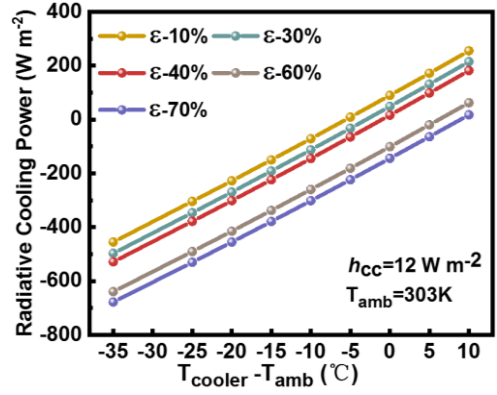

**Fig. S12.** Theoretical cooling/heating power of Es-TPU membrane under different strains (10%, 30%, 40%, 60%, 70%) ($h_{cc}$=12 Wm$^{-2}$, $T_{amb}$=303 K).

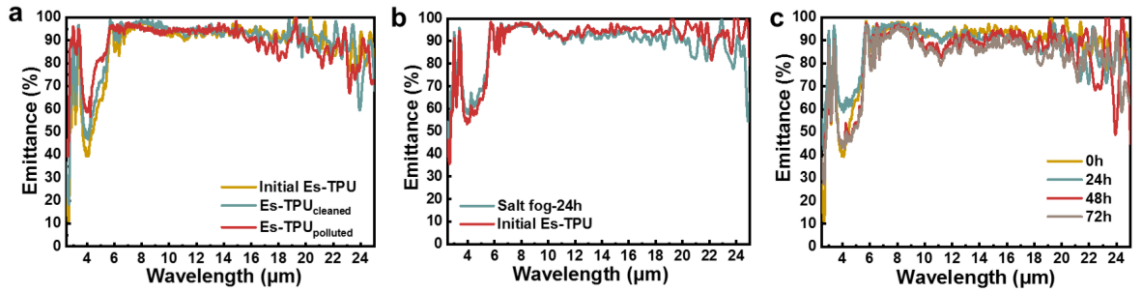

**Fig. S13.** IR emittance of Es-TPU membranes a) initial, polluted and cleaned Es-TPU membranes, b) before and after salt spray treatment for 24 h, c) before and after 0 h, 24 h, 48 h, 72 h of UV aging test.

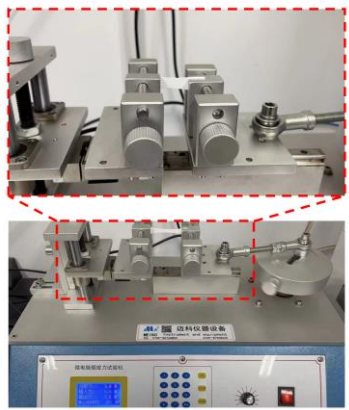

**Fig. S14.** Tensile cycles test device.

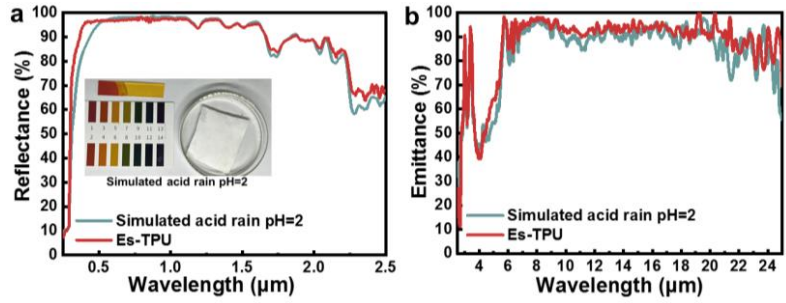

**Fig. S15.** Weatherability of the Es-TPU membrane: a) UV–vis–NIR reflectance, b) IR emittance of Es-TPU membranes before and after simulated acid rain treatment.

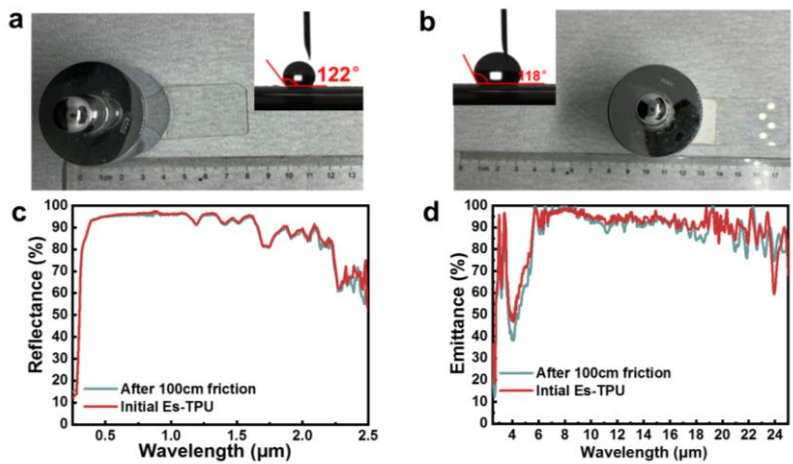

**Fig. S16.** Mechanical durability test of Es-TPU membrane. a-b) Abrasion of 120 mesh sandpaper, inset: contact angle of Es-TPU membranes before and after 100 cm sandpaper abrasion. c) UV–vis–NIR reflectance, b) IR emittance of Es-TPU membranes before and after 100 cm sandpaper abrasion.

We coated the glass house model with Es-TPU-80% and Es-TPU-0%, respectively, to observe the light transmission in a dark environment. A beam of light shines directly on the transparent roof, and it can penetrate directly and form uneven dazzling light spots indoors. The light shines on the top covered with Es-TPU-80%, the spot area on the roof surface becomes more extensive, which can illuminate the room evenly (the spot in the room is evenly distributed on the school badge), and due to the reflection of the film, the amount of light reflected into the surrounding environment increases (the area of surrounding black shadows decreases). When Es-TPU-0% is irradiated, more light is reflected into the surrounding environment, and basically, no light enters the room.

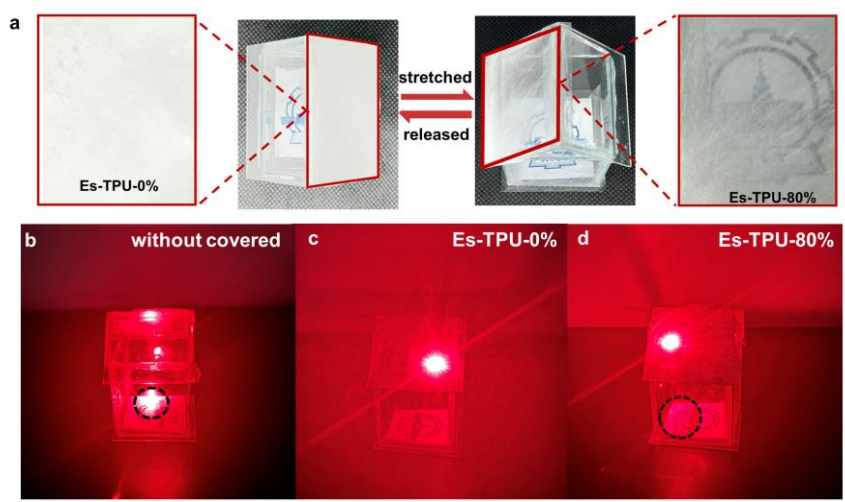

**Fig. S17.** Photographs of a) the Es-TPU membrane coated on the roof of the simulated glass house, light shines on the b) glass roof, d) Es-TPU-0%, and c) Es-TPU-80%.

**Table S1.** The theoretical maximum temperature difference, cooling and heating powers of Es-TPU membrane under different strains on black substrate ($h_{cc}$ is 12 W m$^{-2}$ K$^{-1}$, 303K).

| Strain | $\Delta T_{MAX}$ (°C) | $P_{cool}$ (W m$^{-2}$) | Strain | $\Delta T_{MAX}$ (°C) | $P_{heat}$ (W m$^{-2}$) |
| --- | --- | --- | --- | --- | --- |
| 0% | -7.5 | 118.25 | 50% | 2.8 | 45.25 |
| 10% | -5.8 | 89.68 | 60% | 6.4 | 101.26 |
| 20% | -4.3 | 66.93 | 70% | 9.3 | 144.40 |
| 30% | -3.1 | 48.63 | 80% | 14.3 | 220.34 |
| 40% | -1.1 | 16.12 | | | |